\documentclass[pre,aps,twocolumn,raggedbottom,superscriptaddress,nofootinbib]{revtex4-2}
\pdfoutput=1
\usepackage{amsmath,amssymb,bm,graphicx}
\usepackage[colorlinks=true,citecolor=blue,linkcolor=blue,urlcolor=blue]{hyperref}
\newcommand{\dd}{\mathrm{d}}

\DeclareMathOperator{\Li}{Li}
\begin{document}
\title{Exact nonstationary heat-excess statistics in the weakly asymmetric Kipnis--Marchioro--Presutti model}
\author{Baruch Meerson}
\email{meerson@mail.huji.ac.il}
\affiliation{Racah Institute of Physics, Hebrew University of Jerusalem,
Jerusalem 91904, Israel}
\date{November 19, 2026}
\begin{abstract}
Exact nonstationary large-deviation statistics remain scarce for interacting many-body
systems. Here we employ the macroscopic fluctuation theory (MFT) to determine the complete long-time statistics of heat excess
produced by an initially localized heat pulse in the weakly asymmetric
Kipnis--Marchioro--Presutti model. We find a remarkable nonlocal canonical
transformation which maps the weakly asymmetric MFT problem, including
its initial and terminal conditions, onto the symmetric problem, previously solved exactly.
The resulting rate function is an explicit nonlinear deformation of its
symmetric counterpart. We also extend the solution to heat measured beyond an arbitrary observation point.
\end{abstract}
\maketitle

Rare fluctuations of particle and energy currents reveal features of
nonequilibrium transport that are invisible in the average evolution.
Macroscopic fluctuation theory (MFT) provides a powerful unified framework for their
description in diffusive interacting systems~\cite{MFT}. It expresses the
probability of an atypical transfer through an optimal history of the
density and current. Finding this history, however, usually requires
solving two coupled nonlinear partial differential equations with boundary
conditions imposed at different times. For a nonstationary initial state,
even the deterministic reference evolution depends on time. Exact results
for the complete large-deviation statistics are therefore especially
valuable: they connect typical fluctuations and rare events within one
solution and provide benchmarks for perturbative theories and simulations.

An earlier exact result for nonstationary current statistics was obtained
by Derrida and Gerschenfeld for the microscopic symmetric exclusion
process using Bethe ansatz~\cite{DG2009}. More recently, applications of
the inverse scattering method (ISM) have enabled exact solutions directly
within MFT and related weak-noise theories, including the weak-noise
theory of the Kardar--Parisi--Zhang equation~\cite{KLD2021}.
Bettelheim, Smith, and Meerson (BSM) determined the full long-time
statistics of heat transfer from a localized pulse in the symmetric
Kipnis--Marchioro--Presutti (KMP) model~\cite{BSM,BSMJSTAT}.
Mallick, Moriya, and Sasamoto solved the time-dependent MFT equations
describing the statistics of the integrated current in the symmetric
exclusion process~\cite{MMS}. Subsequently, Krajenbrink and Le Doussal
obtained exact current and tracer statistics for the weakly asymmetric
exclusion process and established MFT integrability for a wider class
of driven models with quadratic mobility~\cite{KLD}. Yet integrability
of the evolution equations alone does not solve a prescribed large-deviation
problem: the initial and terminal conditions must also allow the
scattering data to be determined.

Here we obtain the complete heat-excess statistics for a localized pulse
in the weakly asymmetric KMP model. The key to our solution is a nonlocal canonical
transformation that removes the drive from the MFT equations while
preserving the action and the form of both temporal boundary conditions.
This transformation reduces the driven problem to the symmetric problem 
solved by BSM \cite{BSM}.
Consequently, the entire effect of the drive on the rate function is
encoded in an explicit nonlinear change of its argument. Furthermore, the same
transformation also gives the statistics of the heat beyond a nonzero
observation point, previously determined for the symmetric KMP model 
in Ref. \cite{BSMJSTAT}.

\emph{Model and observable.}
The KMP model~\cite{KMP} describes stochastic exchange of nonnegative
energies between neighboring lattice sites. In its symmetric version, a
pair redistributes its combined energy uniformly. We introduce a bias by
multiplying the redistribution rate density by $e^{\nu Q}$, where $Q$ is
the energy transferred to the right~\cite{DG2017}. The unbiased bond rate
is chosen to be 2, so that the diffusivity is unity. In the 1D weakly
asymmetric hydrodynamic description, valid at long times, the energy density $u(x,t)$ 
obeys the Langevin equation
\begin{equation}
 \partial_t u=\partial_x\left(\partial_x u-2\nu u^2
                         -\sqrt{2}\,u\eta\right),
 \label{eq:fhd}
\end{equation}
where $\eta$ is Gaussian white noise of unit strength. Thus the mobility
is $\sigma(u)=2u^2$, and the deterministic current is $-u_x+2\nu u^2$
\cite{BGL2005,DG2017}. The parameter $\nu$ has dimensions of inverse energy.
The weak-asymmetry description requires $|\nu|u_{\mathrm{loc}}\ll1$,
where $u_{\mathrm{loc}}$ is the characteristic energy per site.

Suppose that initially a heat pulse of total energy $W$ is released at the origin,
$u(x,0)=W\delta(x)$. At time $T\gg1$, we measure the heat excess
\begin{equation}
 J=\int_0^\infty u(x,T)\,\dd x-\frac W2,\qquad j=\frac JW.
 \label{eq:observable}
\end{equation}
The reference is an equal split of the heat, so the typical $j$ need not
vanish in the driven system. Rescaling $x$ by the diffusion length $\sqrt T$, $t$ by $T$, and
$u$ by $W/\sqrt T$ leaves one drive parameter, $\epsilon=\nu W$, and the long-time
probability density of $J$ has the large-deviation form
\begin{equation}
 -\ln\mathcal P(J,T;W,\nu)\simeq\sqrt T\,s(j;\epsilon).
 \label{eq:ldp}
\end{equation}
For the spreading pulse, $u_{\mathrm{loc}}\sim W/\sqrt T$, so the
weak-asymmetry condition becomes $|\epsilon|/\sqrt T\ll1$.
Thus $\epsilon$ need not be small. We assume that the initial microscopic
transient does not affect the leading long-time large-deviation statistics.
Indeed, already over a relatively short timescale (much shorter than $T$), the initial delta function gets ``smeared" over a relatively large region of space, thus justifying the weakly asymmetric condition, but leaving the heat localized from a macroscopic point of view.

\emph{MFT formulation.}
As one can show (the derivation is almost identical to that of Ref. \cite{BSM} for the undriven case $\epsilon=0$), 
the optimal density $u$ and the gradient
$v=-\partial_xp$ of the conjugate field satisfy, in the rescaled variables,  the coupled nonlinear partial differential equations
\begin{align}
 u_t&=\partial_x[u_x+2u^2(v-\epsilon)],\label{eq:u}\\
 v_t&=\partial_x[-v_x+2uv(v-2\epsilon)],\label{eq:v}
\end{align}
with the initial and terminal conditions
\begin{equation}
 u(x,0)=\delta(x),\qquad v(x,1)=-\lambda\delta(x).
 \label{eq:bc}
\end{equation}
The Lagrange multiplier $\lambda$ fixes $\int_0^\infty u(x,1)\dd x=1/2+j$.
The rescaled action and the Legendre relation are
\begin{equation}
 s=\int_0^1\dd t\int_{-\infty}^{\infty}\dd x\,u^2v^2,
 \qquad \partial_j s=\lambda.
 \label{eq:action}
\end{equation}

\emph{Canonical transformation.}
The cumulative heat $h(x,t)=\int_{-\infty}^xu(y,t)\dd y$ provides the
natural variable for removing the drive. Indeed, the deterministic limit of the equation
\[
h_t=h_{xx}+2h_x^2(v-\epsilon)
\]
has the familiar Cole--Hopf structure \cite{Hopf,Cole}. Nonlocal transformations have previously proved useful in relating
nonequilibrium and equilibrium fluctuation problems~\cite{TKL2008}.
Inspired by the exponential cumulative-energy
mappings between asymmetric and symmetric Brownian energy
processes~\cite{CGRS2016,CCF2024}, we introduce a transformed cumulative
heat $H$ and its density $U=\partial_x H$, together with the transformed
conjugate field $V$:
\begin{equation}
 \begin{gathered}
 H=\frac{1-e^{-2\epsilon h}}{1-e^{-2\epsilon}},\\
 U=\frac{\epsilon}{\sinh\epsilon}e^{\epsilon(1-2h)}u,\quad
 V=\frac{\sinh\epsilon}{\epsilon}e^{-\epsilon(1-2h)}v.
 \end{gathered}
 \label{eq:map}
\end{equation}
The expression for $U$ follows by differentiating $H$ with respect
to $x$, using $h_x=u$.
The transformation $(h,v)\mapsto(H,V)$ is canonical:
$V\,\delta H=v\,\delta h$. It preserves positivity and total heat, with
$H(-\infty,t)=0$ and $H(\infty,t)=1$. At $\epsilon=0$, the formulas
are understood by continuity.

Substitution into Eqs.~\eqref{eq:u} and \eqref{eq:v} yields
\begin{equation}
 \begin{aligned}
 U_t&=\partial_x(U_x+2U^2V),\\
 V_t&=\partial_x(-V_x+2UV^2),
 \end{aligned}
 \label{eq:sym}
\end{equation}
precisely the symmetric KMP equations. Also, the action retains its form,
$s=\int_0^1\dd t\int_{-\infty}^{\infty}\dd x\,U^2V^2$. Essentially, this remarkable transformation
extends the deterministic Cole--Hopf reduction to the coupled MFT equations (\ref{eq:u}) and (\ref{eq:v}).

Crucially, the temporal conditions transform just as simply.
The initial step $h(x,0)=\Theta(x)$ maps to $H(x,0)=\Theta(x)$,
so $U(x,0)=\delta(x)$. At the terminal time, $h(0,1)=1/2-j$, and therefore
\begin{equation}
 V(x,1)=-\omega\delta(x),\qquad
 \omega=\lambda\frac{\sinh\epsilon}{\epsilon}e^{-2\epsilon j}.
 \label{eq:omega}
\end{equation}
In its turn, the transformed heat excess $q=\int_0^\infty U(x,1)\dd x-1/2$ is
\begin{equation}
 q=\frac{e^{2\epsilon j}-\cosh\epsilon}{2\sinh\epsilon}.
 \label{eq:q}
\end{equation}
Thus not only the bulk
equations, but the whole boundary-value problem reduces to the symmetric one. The map $j \to q$ is monotone and transforms
$j=\pm1/2$ to $q=\pm1/2$.

\emph{Exact heat-excess statistics.}
Let $s_0(q)$ be the symmetric rate function, which corresponds to $\epsilon=0$ and was obtained by BSM~\cite{BSM}.
The action is invariant under the canonical transformation,
yielding our main result:
\begin{equation}
s(j;\epsilon)=s_0\!\left(
 \frac{e^{2\epsilon j}-\cosh\epsilon}{2\sinh\epsilon}\right).
 \label{eq:rate}
\end{equation}
The transformation also preserves the cost of arbitrary density-current
histories, so this relation identifies the constrained minima of the two
MFT problems (see Appendix~\ref{app:canonical}). The result covers the full interval $-1/2<j<1/2$.
A simple sanity check is provided by the limit of $\epsilon \to 0$: $s(j;\epsilon \to 0) = s_0(j)$.

For an explicit evaluation, the BSM solution \cite{BSM} can be written as
\begin{align}
 j_0(\omega)&=\frac{1}{4\pi\omega}\int_{-\infty}^{\infty}\frac{\dd k}{k^2}
              \ln(1+\omega^2k^2e^{-2k^2}),\label{eq:j0}\\
 s_0(j_0(\omega))&=\omega j_0(\omega)
 +\frac{1}{8\pi}\int_{-\infty}^{\infty}\frac{\dd k}{k^2}
                 \Li_2(-\omega^2k^2e^{-2k^2}).\label{eq:s0}
\end{align}
Here $|\omega|<\infty$, $\Li_2(z)=-\int_0^z t^{-1}\ln(1-t)\,\dd t$ for $z\leq0$ is the dilogarithm function, and $j_0(0)=0$. The action of the driven system has
the same parametric value, while the heat excess becomes
\begin{equation}
 j(\omega;\epsilon)=\frac{\ln[\cosh\epsilon+
                         2\sinh\epsilon\,j_0(\omega)]}{2\epsilon}.
 \label{eq:main}
\end{equation}
Equations~\eqref{eq:s0} and \eqref{eq:main} give the full rate function
without reconstructing the optimal history. The Legendre identity is
preserved: $\partial_js=\omega\,\partial_jq=\lambda$.

\begin{figure}[t]
 \centering
 \includegraphics[width=\columnwidth]{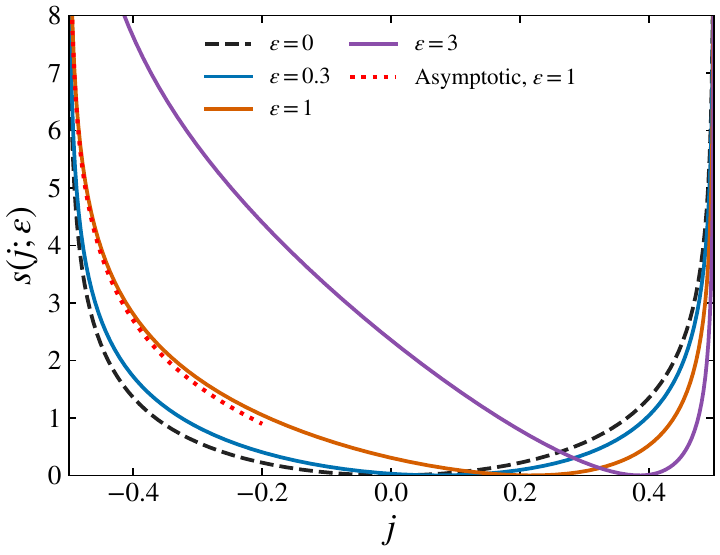}
 \caption{\label{fig:action}Rescaled action of the heat excess for
 $\epsilon=0$ (black dashed), $0.3$ (blue), $1$ (orange), and $3$ (purple),
 obtained from Eqs.~\eqref{eq:s0} and \eqref{eq:main}.
 The drive shifts the zero-action minimum to
 $\bar j=\ln(\cosh\epsilon)/(2\epsilon)$ and makes the two tails
 asymmetric. The action diverges at $j=\pm1/2$; the vertical range is
 truncated at $s=8$. The red dotted curve shows the parametric
 left-endpoint asymptotic for $\epsilon=1$, given by
 Eqs.~\eqref{eq:paramcurrent} and \eqref{eq:paramaction}.}
\end{figure}

Figure~\ref{fig:action} shows how the drive $\epsilon$ deforms the symmetric result.
The minimum  of the rate function moves toward positive heat excess, and the two tails become
asymmetric. Positive driving makes it less costly to transfer almost all
the heat to the right and more costly to transfer it to the left.
Both endpoint costs still diverge. Reversing the drive reflects the rate
function, $s(j;\epsilon)=s(-j;-\epsilon)$, as expected on physical grounds.

\emph{Typical fluctuations.}
The zero-action value and the curvature at the minimum follow directly
from Eqs.~\eqref{eq:j0} and \eqref{eq:main}:
\begin{equation}
 \bar j=\frac{\ln\cosh\epsilon}{2\epsilon},\qquad
 \chi(\epsilon)=\frac{1}{4\sqrt{2\pi}}
                    \left(\frac{\tanh\epsilon}{\epsilon}\right)^2,
 \label{eq:chi}
\end{equation}
where $s\simeq(j-\bar j)^2/(2\chi)$ and
$\operatorname{Var}(j)\simeq\chi/\sqrt T$. The typical value also follows from the
Cole--Hopf solution of the deterministic Burgers equation. Linearizing
the Langevin equation~(\ref{eq:fhd}) around that solution independently reproduces
the variance (see Appendix~\ref{app:checks}). The drive thus both shifts the typical heat split
and suppresses its fluctuations.

\emph{Endpoint asymptotics.}
For fixed $\epsilon$, let $j=\pm(1/2-\delta)$, with
$\delta\to0^+$, and write $L=\ln(1/\delta)$. The two tails obey
\begin{multline}
 s\!\left(\pm\left(\frac12-\delta\right);\epsilon\right)
 =\frac{4}{3\pi}L^{3/2} +\frac{2\sqrt L}{\pi}\ln L\\
 +\frac{2\sqrt L}{\pi} \left[\ln\!\left(\frac{2\sinh\epsilon}{\pi\epsilon}\right)
 -(1\pm\epsilon)\right]+o(\sqrt L).
 \label{eq:endpoint}
\end{multline}
The upper and lower signs correspond to the right and left endpoints,
respectively; the $\epsilon=0$ limit is understood by continuity.
As we see, the leading divergence is independent of the drive. Its first effect
appears at order $\sqrt L$, below the universal $\sqrt L\ln L$
correction. In particular,
\begin{equation}
 \begin{aligned}
 &s\!\left(-\frac12+\delta;\epsilon\right)
 -s\!\left(\frac12-\delta;\epsilon\right)\\
 &\hspace{1cm}=\frac{4\epsilon}{\pi}\sqrt{\ln(1/\delta)}
 +o\!\left(\sqrt{\ln(1/\delta)}\right).
 \end{aligned}
 \label{eq:taildifference}
\end{equation}
The difference between the two actions diverges, although their
ratio tends to one. Appendix~\ref{app:tails} derives these asymptotic results from
the symmetric BSM solution.

The logarithmic expansion~\eqref{eq:endpoint} converges very slowly, so its
accuracy remains limited over the range of actions shown in
Fig.~\ref{fig:action}. We therefore plotted, for $\epsilon=1$, a
parametric asymptotic (the red dotted curve) that retains the large-multiplier expansion and
the exact transformation of the endpoint distance, without re-expanding
in $L$. This gives substantially better agreement with the exact
$\epsilon=1$ curve, see Fig.~\ref{fig:action}. The parametric formulas are given in
Appendix~\ref{app:tails}.

\emph{Nonzero observation point.}
The reduction extends to the heat fraction
$m=W^{-1}\int_X^\infty u(x,T)\dd x$, with $\xi=X/\sqrt T$.
The terminal delta function is now at $x=\xi$ \cite{BSMJSTAT}, and its position is unchanged
by Eq.~\eqref{eq:map}. The transformed heat fraction is
$m_0=(e^{2\epsilon m}-1)/(e^{2\epsilon}-1)$. Hence
\begin{equation}
 s_{\rm WAKMP}(m,\xi;\epsilon)=s_{\rm BSM}\!\left(
             \frac{e^{2\epsilon m}-1}{e^{2\epsilon}-1},\xi\right),
 \label{eq:finiteX}
\end{equation}
where $s_{\rm BSM}$ is the full symmetric rate function derived in
Ref.~\cite{BSMJSTAT}. At $X=0$, this reduces to Eq.~\eqref{eq:rate}
with $m=1/2+j$. Thus a single transformation solves an entire family
of nonstationary heat-transfer problems parametrized by $X$; details are given in
Appendix~\ref{app:finiteX}.

\emph{Discussion.}
We have obtained an exact large-deviation function for nonstationary
heat transfer in the weakly asymmetric KMP model. 
The result determines the full
statistics throughout the allowed range of heat excess, connecting the
Gaussian center to the strongly non-Gaussian tails. The simplicity of the result
comes
from a nonlocal canonical transformation that absorbs the drive into the cumulative
heat and the terminal constraint. The microscopic dynamics and the
optimal profiles change with the drive, but their MFT cost is inherited
from the symmetric problem.

The decisive feature is the compatibility of the transformation with the localized initial pulse and the terminal heat measurement. This compatibility allows us to map the entire fluctuation problem onto the exactly solved symmetric problem. In its turn, the quadratic mobility of the KMP model ensures that the reciprocal rescaling of the density $u$ and field $v$ preserves the action density, \(U^2V^2=u^2v^2\). Consequently, the driven rate function follows directly from its symmetric counterpart, without reconstructing the optimal history.  Thus, a single canonical transformation yields an entire family of exact nonstationary large-deviation functions for driven heat transport, with arbitrary drive parameter \(\epsilon\) and observation point.

\textit{Acknowledgments}. The author thanks Naftali R. Smith for useful comments. This research was supported by the Israel
Science Foundation (Grant No. 1579/25). The author acknowledges the use of
ChatGPT (OpenAI, GPT-6 series) to assist in derivations, plotting and manuscript drafting. The AI output was reviewed and
verified by the author, who takes full responsibility for the results presented in this work.

\emph{Data availability.}
The analytical results supporting this study are contained in the article.
Figure~\ref{fig:action} is obtained by numerical quadrature of
Eqs.~\eqref{eq:j0} and \eqref{eq:s0}, followed by the transformation
in Eq.~\eqref{eq:main}; these expressions and Eqs. (\ref{eq:paramcurrent}) and (\ref{eq:paramaction}) specify the plotted curves
completely. No external datasets were used.

\appendix

\section{Canonical reduction and invariance of the action}\label{app:canonical}
Using $H(h)$ from Eq.~\eqref{eq:map}, set
\begin{equation}
\begin{gathered}
 a(h)=\frac{\partial H}{\partial h}=\frac{\epsilon}{\sinh\epsilon}e^{\epsilon(1-2h)},\\
  U=au,\\ V=v/a.
\end{gathered}
\end{equation}
Since $V\,\delta H=v\,\delta h$, we have $V H_t=v h_t$.
Thus the transformation preserves the time-derivative term
in the Hamiltonian action and is therefore canonical.
To verify the transformed evolution equations, use
$h_t=u_x+2u^2(v-\epsilon)$.
Then $a_x=-2\epsilon ua$ and
$a_t=-2\epsilon h_ta$, so
\begin{equation}
\begin{gathered}
 U_t=\partial_x(ah_t),\\ U_x+2U^2V=ah_t.
\end{gathered}
\end{equation}
For the second equation,
\begin{align}
 V_t&=\frac{v_t+2\epsilon h_tv}{a},\\
 V_{xx}&=\frac{v_{xx}+2\epsilon u_xv+4\epsilon uv_x+4\epsilon^2u^2v}{a},\\
 2\partial_x(UV^2)&=\frac{2\partial_x(uv^2)+4\epsilon u^2v^2}{a}.
\end{align}
Using Eq.~\eqref{eq:v} gives $V_t=-V_{xx}+2\partial_x(UV^2)$.
The initial pulse is transformed through its cumulative step:
$h(x,0)=\Theta(x)$ maps to $H(x,0)=\Theta(x)$,
hence $U(x,0)=\delta(x)$.
Equivalently, one may first use a narrow, nonnegative unit-mass pulse
and then let its width vanish.

For an arbitrary
density-current history obeying $u_t+\partial_x\mathcal J=0$, its MFT action cost is
\begin{equation}
 \mathcal S_\epsilon[u,\mathcal J]=\int_0^1\dd t\int_{-\infty}^{\infty}\dd x\,
 \frac{(\mathcal J+u_x-2\epsilon u^2)^2}{4u^2}.
\end{equation}
With vanishing flux at $-\infty$, we obtain $h_t=-\mathcal J$. Let us define the transformed
current $\mathcal K=a\mathcal J$. Then $H_t=-\mathcal K$ and
$U_t+\partial_x\mathcal K=0$, while
\begin{equation}
\begin{gathered}
 \mathcal K+U_x=a(\mathcal J+u_x-2\epsilon u^2),\\
 \mathcal S_0[U,\mathcal K]=\mathcal S_\epsilon[u,\mathcal J].
\end{gathered}
\end{equation}
The cumulative map is positive and invertible on $0\le h\le1$:
\begin{equation}
 h=-\frac{1}{2\epsilon}\ln[1-(1-e^{-2\epsilon})H].
\end{equation}
It maps the terminal heat constraints according to Eq.~\eqref{eq:q}.
Minimization therefore gives Eq.~\eqref{eq:rate}. 

\section{Independent checks: deterministic evolution and variance}\label{app:checks}
At $\lambda=0$ we have $v=0$, and the density obeys the deterministic Burgers
equation $\bar u_t=\bar u_{xx}-2\epsilon\partial_x(\bar u^2)$.
Its cumulative heat satisfies the equation
$\bar h_t=\bar h_{xx}-2\epsilon\bar h_x^2$. The Cole--Hopf field
$Z(x,t)=e^{-2\epsilon\bar h(x,t)}$ solves the diffusion equation, with 
with the step initial condition $Z(x,0)=1$ for $x<0$
and $Z(x,0)=e^{-2\epsilon}$ for $x>0$. Define
\begin{equation}
 G(x,t)=\frac{e^{-x^2/(4t)}}{\sqrt{4\pi t}}.
\end{equation}
Then
\begin{equation}
\begin{gathered}
 Z(x,t)=1-\frac{1-e^{-2\epsilon}}{2}
 \left[1+\operatorname{erf}\left(\frac{x}{\sqrt{4t}}\right)\right],\\
 \bar u(x,t)=\frac{1-e^{-2\epsilon}}{2\epsilon}\frac{G(x,t)}{Z(x,t)}.
\end{gathered}
\end{equation}
At the origin, $Z(0,1)=e^{-\epsilon}\cosh\epsilon$ and
$\bar j=1/2-\bar h(0,1)=\ln(\cosh\epsilon)/(2\epsilon)$,
in agreement with Eq.~\eqref{eq:chi}.

To calculate the variance directly from the Langevin equation (\ref{eq:fhd}), we expand
$h=\bar h+T^{-1/4}h_1+\cdots$, exploiting the small parameter $T^{-1/4}$.  The rescaled current noise is
$\sqrt2T^{-1/4}u\eta$, and to linear order
\begin{equation}
\begin{gathered}
 (h_1)_t=(h_1)_{xx}-4\epsilon\bar u(h_1)_x-\sqrt2\bar u\eta,\\
  h_1(x,0)=0.
\end{gathered}
\end{equation}
Using $Z_t=Z_{xx}$ and $Z_x=-2\epsilon\bar uZ$, one finds that
multiplication by $Z$ cancels the drift term, yielding
$\partial_t(Zh_1)=\partial_x^2(Zh_1)-\sqrt2Z\bar u\eta$, where
\begin{equation}
\quad Z\bar u=\frac{1-e^{-2\epsilon}}{2\epsilon}G(x,t) 
\end{equation}
Since the heat excess is $j=1/2-h(0,1)$, its fluctuation is
$j-\bar j=-T^{-1/4}h_1(0,1)$. Evaluating the heat-kernel
solution at the observation point $x=0$ and terminal time $t=1$ gives

\begin{equation}
 \begin{aligned}
 h_1(0,1)={}&-\sqrt2\,\frac{\tanh\epsilon}{\epsilon}\\
 &\times\int_0^1\dd t\int_{-\infty}^{\infty}\dd x\,G(x,1-t)G(x,t)\eta(x,t).
 \end{aligned}
\end{equation}
Evaluating the double integral,
\begin{equation}
 \int_0^1\dd t\int_{-\infty}^{\infty}\dd x\,G(x,1-t)^2G(x,t)^2
 =\frac{1}{8\sqrt{2\pi}},
\end{equation}
we recover the variance in Eq.~\eqref{eq:chi}.

\section{Endpoint asymptotics}\label{app:tails}
We first refine the symmetric endpoint asymptotic to the order needed
to resolve the effect of the drive. By symmetry, take $\omega>0$ and
write $\mu=\ln\omega$. Equation (S24) of the Supplemental Material
to Ref.~\cite{BSM}, together with Eq.~\eqref{eq:j0}, gives
\begin{equation}
 \begin{aligned}
 D(\mu)&\equiv e^\mu  \left(\frac12-j_0(e^\mu)\right)\\
 &=\frac{2\sqrt\mu}{\pi}
 +\frac{\ln\mu+2}{2\pi\sqrt\mu}+o(\mu^{-1/2}).
 \end{aligned}
 \label{eq:tailD}
\end{equation}
The Legendre identity yields
\[
\frac{\dd s_0[j_0(e^\mu)]}{\dd\mu}=D-D'.
\]
Integrating over $\mu$,
we obtain
\begin{equation}
 s_0[j_0(e^\mu)]=
 \frac{4}{3\pi}\mu^{3/2}
 +\frac{\sqrt\mu}{\pi}(\ln\mu-2)+o(\sqrt\mu).
 \label{eq:tailsmu}
\end{equation}
Defining $L_0=\ln[1/(1/2-|q|)]$, the relation
$1/2-|q|=e^{-\mu}D(\mu)$ yields
\begin{equation}
 \mu=L_0+\frac12\ln L_0+\ln\frac2\pi+o(1).
 \label{eq:tailinvert}
\end{equation}
Substituting into Eq.~\eqref{eq:tailsmu} gives
\begin{equation}
 \begin{aligned}
 s_0(q)={}&\frac{4}{3\pi}L_0^{3/2}\\
 &+\frac{2\sqrt{L_0}}{\pi}
 \left(\ln L_0+\ln\frac2\pi-1\right)+o(\sqrt{L_0}).
 \end{aligned}
 \label{eq:symmetricendpoint}
\end{equation}

We now use the exact mapping~\eqref{eq:q}. For
$j=\pm(1/2-\delta)$ and fixed $\epsilon$, it gives
\begin{equation}
 \frac12-|q|=\frac{\epsilon e^{\pm\epsilon}}{\sinh\epsilon}\,
 \delta+O(\delta^2),
 \label{eq:endpointdistance}
\end{equation}
where $q$ approaches the endpoint with the same sign as $j$.
Consequently, with $L=\ln(1/\delta)$,
\begin{equation}
 L_0=L-\ln\!\left(\frac{\epsilon}{\sinh\epsilon}\right)
 \mp\epsilon+O(\delta).
 \label{eq:endpointlog}
\end{equation}
Inserting this expression into Eq.~\eqref{eq:symmetricendpoint}
and using $s(j;\epsilon)=s_0(q)$ yields Eq.~\eqref{eq:endpoint}.
Subtracting the right-endpoint expansion from the left-endpoint
expansion gives Eq.~\eqref{eq:taildifference}.

\subsection*{Parametric asymptotic for comparison with the exact result}
The explicit expansion~\eqref{eq:endpoint} makes the drive dependence
transparent, but converges very slowly. In particular, expanding
$\mu^{3/2}$ after Eq.~\eqref{eq:tailinvert} generates omitted terms
of order $(\ln L_0)^2/\sqrt{L_0}$, which remain appreciable at
moderate $L_0$. Expanding the finite shift between $L_0$ and $L$
introduces further corrections.

For numerical comparison it is more accurate to retain $\mu$ as the
parameter.
At the left endpoint, $j=-1/2+\delta$, Eq.~\eqref{eq:q} gives
$1/2-|q|=(e^{2\epsilon\delta}-1)/(e^{2\epsilon}-1)$ exactly.
Keeping this relation unexpanded and using the truncated
expansion~\eqref{eq:tailD} yields the parametric approximation
\begin{equation}
 \begin{aligned}
 j_{\rm as}(\mu;\epsilon)={}&-\frac{1}{2} +\frac{1}{2\epsilon}\ln\Biggl[1+(e^{2\epsilon}-1)e^{-\mu}\\
 &\qquad\quad\times\left(\frac{2\sqrt\mu}{\pi}
 +\frac{\ln\mu+2}{2\pi\sqrt\mu}\right)\Biggr],
 \end{aligned}
 \label{eq:paramcurrent}
\end{equation}
\begin{equation}
 s_{\rm as}(\mu)=\frac{4}{3\pi}\mu^{3/2}
 +\frac{\sqrt\mu}{\pi}(\ln\mu-2),
 \label{eq:paramaction}
\end{equation}
where $\mu\gg1$ and the $\epsilon=0$ limit is taken by continuity.
These asymptotic formulas use the same large-multiplier terms as
Eq.~\eqref{eq:endpoint}, but avoid the additional logarithmic
expansions. They produce the red dotted curve in Fig.~\ref{fig:action}.

\section{Heat beyond a nonzero observation point}\label{app:finiteX}
At $x=\xi$, the terminal cumulative heat is $h(\xi,1)=1-m$.
Consequently,
\begin{equation}
\begin{gathered}
 m_0=1-H(\xi,1)=\frac{e^{2\epsilon m}-1}{e^{2\epsilon}-1},\\
 V(x,1)=-\omega\delta(x-\xi),\\
 \omega=\lambda\frac{\sinh\epsilon}{\epsilon}e^{\epsilon(1-2m)}.
\end{gathered}
\end{equation}
If the symmetric solution is parameterized by $m_0(\omega,\xi)$,
the driven solution is
\begin{align}
 m(\omega,\xi;\epsilon)&=\frac{1}{2\epsilon}
 \ln[1+(e^{2\epsilon}-1)m_0(\omega,\xi)],\\
 \lambda(\omega,\xi;\epsilon)&=
 \frac{2\epsilon\omega}{e^{2\epsilon}-1}
 [1+(e^{2\epsilon}-1)m_0(\omega,\xi)].
\end{align}
The full symmetric solution, with its physical branch continuations,
is taken from Ref.~\cite{BSMJSTAT}. Its action is unchanged.
The deterministic fraction follows by substituting
$\bar m_0=\tfrac12\operatorname{erfc}(\xi/2)$:
\begin{equation}
 \bar m(\xi;\epsilon)=\frac{1}{2\epsilon}
 \ln\left[1+\frac{e^{2\epsilon}-1}{2}\operatorname{erfc}\left(\frac\xi2\right)\right].
\end{equation}
This result also follows from the Cole--Hopf solution, described above.

\end{document}